\documentclass[aps,preprint,showpacs,floatfix]{revtex4}

\usepackage{mathptmx}
\usepackage{courier}

\usepackage{amsmath}
\usepackage{amssymb}

\usepackage{graphicx}

\begin{document}
  
  \title{Microsolvation of Li$^+$ in small He Clusters. Li$^+$He$_n$ species from 
    Classical and Quantum Calculations}
  \author{C. Di Paola} 
  \author{F. Sebastianelli}
  \author{E. Bodo}
  \author{I. Baccarelli}
  \author{F.A. Gianturco\footnote{corresponding author. e-mail:
      fa.gianturco@caspur.it  fax: +39-06-49913305}}
  
  \affiliation{Department of Chemistry, University of Rome ``La Sapienza''and INFM, Piazzale
    A. Moro 5, 00185 Rome, Italy}
  
  \author{M. Yurtsever}
  \affiliation{Istanbul Technical University, Chemistry Department, 80626 Maslak, Istanbul, Turkey}

\begin{abstract}
A structural study of the smaller Li$^+$He$_n$ clusters with $n\le30$ has been carried
out using different theoretical methods. The structures and the
energetics of the clusters have been obtained using both classical
energy minimization methods and quantum Diffusion Monte Carlo. The total
interaction acting within the clusters has been obtained as a sum of
pairwise potentials: Li$^+$-He and He-He. This approximation had been
shown in our earlier study \cite{8} to give substantially correct results for energies and geometries
once compared to full ab-initio calculations. 
The general features of the spatial structures, and their
energetics,  are discussed in details for the clusters up to $n=30$
and the first solvation shell is shown to be essentially completed by
the first  ten helium atoms.
\end{abstract}

\pacs{34.20.-b,34.30.+h}

\maketitle

\section{Introduction}

The study of the nanoscopic forces which act within small
aggregates of weakly interacting particles, especially within
assemblies containing helium atoms, has received a great deal of
attention in the last few years \cite{1,2,3} because of the broad
range of phenomena that can be probed under the very
special conditions provided by the He nanodroplets as containers
of atoms or molecules. They are indeed, ultracold homogeneous matrices
where the corresponding spectra often reach very high resolution
due to the superfluid properties of the helium droplet \cite{4}.

The possibility of causing electronic excitation and/or ionization
of the dopant species can offer additional ways of probing the
modified interaction between the new ionic species and the gentle
matrix of the helium droplets \cite{3} since the ensuing distribution
surrounding the molecular impurity is usually markedly deformed as a
consequence of the electrostriction effects on the solvent brought on
by the induction forces between the charged dopant and the helium
atoms in the droplet \cite{5a,5}. The analysis of such effects in the
case of lithium-containing impurities has been particularly intriguing
because of the expected simplicity of the electronic structures
involved and, at the same time, the unusual bonding behaviour of such
systems.  Furthermore, experiments in which the droplet was ionized
after capture of Li atoms \cite{7} have revealed a wealth of newly
formed species like Li$^+$, Li$^+_2$ and LiHe$^+$ which are produced
during droplet fragmentation and evaporation thereby triggering 
the corresponding analysis of structure and
bonding behaviour from theory and computations.

In a previous computational investigation \cite{8} on the
interaction of Li$^+$ with $n$ helium atoms, with $n$ varying from 1 to 6, we
have shown, in fact, that the energy optimized structures of
the clusters were largely determined by two-body (2B)
forces, with the many-body effects being fairly negligible for determining the
final geometry of those small aggregates. In the
present study we have therefore decided to analyze in greater detail
the quantum states and binding energies of the larger structures of
Li$^+$ impurities within the helium droplets in order to extend our 
computational knowledge on such species.

The next Section briefly describes the LiHe$^+$ Potential Energy Curve
(PEC) used here. The results for the trimer ground state are given and
discussed in Section III, where we report the features  obtained using the
distributed Gaussian functions (DGF) method \cite{11,12,13,14} and
the quantum Diffusion Monte Carlo (DMC) method \cite{9}, in comparison
with the values obtained {\it via} a classical minimization procedure.
Section IV extends the discussion to the larger species, whose energetics and
structural features, both obtained with DMC and classical methods, are analyzed.

\section{The pairwise potentials}

The actual potential energy curve (PEC) for Li$^+$He has been studied
before because of its interest in
modeling low-energy plasmas \cite{16}. Alrich and coworkers
suggested earlier on a model potential as a variant of the Tang-Toennies
model \cite{17} while recent calculations of
Sold\'{a}n et al. \cite{19} employed a CCSD(T) treatment
and used the aug-cc-pV5Z basis set. Recently, we have also
carried out a new set of calculations on these clusters \cite{8} using the MP4 method 
and the cc-pV-5Z basis set. The results from
the ab initio calculation of Ref.  \cite{19}, the potential obtained
from a MP4/cc-pV-5Z calculation and  the model potential of
Ref. \cite{17} are presented
in pictorial form by the two panels of Fig. 1 where in the left panel
we show the region of the potential minimum.
One clearly sees there that the deepest well is presented by the
CCSD(T) calculations of \cite{19}, while the model potential of
\cite{17} has the most shallow well. The corresponding long-range
part of the PEC's is reported on the right-hand side panel, where
one can see that the calculations of
\cite{19} closely follow the long-range dipole polarisability
tail, computed here with $\alpha _{He}$=1.3832 $a_0^3$ \cite{20}.
In Table 1 we report the bound states of the Li$^+$-He
diatom when $J=0$ for the three different potentials calculated using a DVR
scheme \cite{dvr} using 2000 grid points in the interval 1.5, 200
a.u.. In the case of the potential of Murrel et al. \cite{19} our results are
almost coincident with those provided by the same authors. 
As can be seen there, the three potentials yield rather
similar energies, although in the well region the agreement is better between
those from the MP4 potential and those from Soldan et al.. On the other hand, the
latter has a long range behavior very similar to that from the semi-empirical
ATT potential (see also Fig. 1 where we compare it to the
``experimental'' long range behavior): both these potentials, indeed, support 8 bound
states while the potential calculated by us with MP4 \cite{8} yields only
7 bound states for $J=0$. The MP4 potential therefore seems not to provide a 
good description of the long range, inductive tail because of the lack of  augmented atomic
functions in the basis set from which it has been obtained. 
Hence, we have decided to employ in the following the CCSD(T) PEC from Soldan
et al.. 

For the He$_2$ dimer we have employed the semiempirical potential (LM2M2)
of Aziz and Slaman \cite{24}: in Fig. 2 we sketch the two pair potentials
used in our calculations. The two potentials are
completely different: the Li$^+$-He one is dominated by
electrostatic and induction forces while the He-He interaction is much weaker
and purely van der Waals in nature. 

As we mentioned above, the full interaction that acts in the M$^+$Rg$_n$ clusters can
be approximated by a sum of pairwise potentials. The
main source of error in this model is due to the absence of the
repulsive interactions between the induced dipoles on the helium
atoms, especially those that are located near the ionic center.
These interactions, however, can be taken to be very small for
rare gas atoms and especially for helium partners. For example, 
in Ref. \cite{29} for the analogous situation of the
anionic dopant Cl$^-$ in Ar clusters it has been shown that the
inclusion of the leading terms of such 3-body forces does not alter
substantially the energetics and the structure of the clusters: in ref. \cite{8}
we have discussed and confirmed this feature for the system analyzed here.
Furthermore, since the charge delocalization in Li$^+$-He$_n$ is very
small \cite{28} and the interaction is mostly determined by
\"{}physical\"{} forces (charge/induced multipole) rather than by
\"{}chemical\"{} effects,  all the helium atoms that are
attached directly to the ionic center can be considered to be
equivalent because there is no detectable tendency for the latter
to preferentially form chemically bound structures with only a subset
of such  adatoms, a fact which would therefore differentiate some of the binding effects with respect to those
in the rest of the cluster.

\section{An outline of the computational tools}

\subsection{The classical optimization}

The total potential in each cluster is described by the sum of
pairwise potentials and searching for the global minimum in this 
hypersurface $V_{TOT}$ will give us the
lowest energy structure for each aggregate from a classical
picture viewpoint. All the classical minimizations were carried out
with a modified version of the OPTIM code by Wales \cite{wales94}. 
This code is based on the eigenvector
following method. The basis of this method is the introduction of
an additional Lagrange parameter into a ``traditional'' optimization
framework \cite{baker86},
which seeks to simultaneously minimize in all directions. 
All searches were conducted in Cartesian
coordinates using projection operators to remove overall translation
and rotation following Baker and Hehre \cite{baker91}. Analytic first
and second derivatives of the energy were employed at every step,
and the resulting stationary point energies and geometries are essentially
exact for the model potential in question. The details of the method
as applied in our group have been given before and therefore will not
be repeated here.

\subsection{The quantum stochastic calculation (DMC)}

The Diffusion MonteCarlo Method has been extensively discussed in a number of papers
(Refs. \cite{hammond,ceperley,suhm} and references therein) and therefore will not be repeated here.

In our implementation a random walk technique is used to solve the
diffusion equation where a large number of random walkers is propagated with time steps
$\Delta\tau$ starting from an arbitrarily chosen initial distribution.
The ground state energy $E_{0}$ is obtained by
averaging $E_{L}(\mathbf{r})$ over the final mixed distributions
$f(\mathbf{r},t)=\psi_{T}(\mathbf{r},t)\psi(\mathbf{r},t)$:
\begin{equation}
<E_{L}>=\frac{\int E_{L}(\mathbf{r})f(\mathbf{r},\tau_{f})d\mathbf{r}}{\int f(\mathbf{r},\tau_{f})d\mathbf{r}}=
\frac{\int\psi_{0}(\mathbf{r})\hat{H}\psi_{T}(\mathbf{r})d\mathbf{r}}{\int\psi_{0}(\mathbf{r})
\psi_{T}(\mathbf{r})d\mathbf{r}}=E_{0}
\end{equation}
The energy is therefore affected by a bias due to the use of a mixed
distribution. The bias is however minimized by using very long propagation
time and very short timesteps.
Expectation values of position operators $\hat{A}(\mathbf{r})$ are
also given by averaging over $f(\mathbf{r})$, a procedure that leads to
biased distributions. However, we believe that the
bias does not modify the qualitative picture of the present calculations
especially because whenever our spatial distributions are compared with those from other
quantum calculations (where possible) and  with 
classically minimized structures, they are found to be in remarkable agreement
with each other as we shall discuss further below.  

The trial function
used here for the $\mathrm{He-He}$ pairs is a product of trial
wavefunction:
\begin{equation}
\Psi_{T}=\prod_{i,j\in\mathrm{He}}\exp\left(-\frac{p_{5}}{R_{ij}^{5}}-
\frac{p_{2}}{R_{ij}^{3}}-p_{0}\log R_{ij}-p_{1}R_{ij}\right)\end{equation}
where the values of the coefficients have been taken by Ref \cite{lewerenz97}.
The trial function for the $\mathrm{Li^{+}-He}$ pair has been chosen
as a gaussian function centered around the energy minimum 
of the relative interaction. However, in order to adjust each
trial wavefunction to the size of the larger clusters, its parameters
have been chosen so that they make it increasingly more delocalized
as  the number of adatoms is increased. All prameters are available
from direct requests to the corresponding author.  

\subsection{The distributed Gaussian Functions (DGF) expansion}

The DGF method \cite{11,12} is a variational approach which solves the
trimer bound state problem written in terms of the atom-atom
coordinates by employing a large set of distributed Gaussian functions
as a basis set. 

First of all, the Hamiltonian for a triatomic system is expressed  
in terms of atom pair coordinates R$_1$, R$_2$ and R$_3$, i.e. in terms of the
distances between each pair of atoms along which the Gaussian functions are distributed
(see \cite{13} for the expression of the Hamiltonian for a triatomic system with an 
impurity). 

The total potential for the trimer is assumed to be the sum of the three two-Body 
potentials as discussed previously and the calculations are carried 
out for a zero total angular momentum.
The total wavefunction for the {\sl v}-th vibrational states is then
expanded in terms of symmetrized basis functions 
\begin{equation}
\label{TOTWF2d}
\Phi _v(R_1,R_2,R_3)=\sum_{j}a_j^{(v)}\phi _j(R_1,R_2,R_3)
\end{equation}
\noindent with 
\begin{equation}
\label{phifun2id}
\phi _j(R_1,R_2,R_3)  =
N_{lmn}^{-1/2}\sum_{P\in S_2}
P[\varphi _l(R_1)\varphi _m(R_2)]\varphi _n(R_3)
\end{equation}
\noindent for the two-identical-particle system. Here, $j$ denotes a
collective index such as 
$j=(l\leq m ; n)$ and N$_{lmn}$ is a normalization constant expressed in term
of the overlap integrals s$_{pq}= \langle \varphi _p |\varphi _q \rangle$.  
Each one-dimensional function $\varphi_p$ is chosen to be a 
DGF (\cite{light86}) centered at the $R_p$ position
\begin{equation}
\label{GAUSS}
\varphi_p (R_i)= \root 4 \of {2 A_p \over \pi}
e^{- A_p(R_i-R_p)^2}.
\end{equation}
With the DGF approach we can obtain several indicators on the spatial
behaviour of the bound states of the systems (the root mean value of the square area,
the average of the cosine value - and the various moments - of any angle
{\sl etc.}), along with several probability distribution functions like
the pair distribution function
\begin{eqnarray}
\label{Dr}
D^{(v)}(R_1) & = & \iint \mid \Phi_v(R_1,R_2,R_3)
\mid ^2 dR_2 dR_3.
\end{eqnarray}
The selection of a suitable set of Gaussian functions and their
distribution within the physical space where the bound states are
located is obviously of primary importance in order to finally obtain
converged and stable results. We extensively experimented with
different sets obtained by changing the number and location of
the DGF depending on the features of the 2B potentials employed. The
details of the basis set employed for the title system are
specifically given in the following section.

\section{The $\mathrm{Li}^+\mathrm{He}_2$ trimer}

Together with the classical optimization and the DMC calculations, we
further carried out the analysis of the properties of the trimer
Li$^{+}$He$_{2}$ using the DGF method.
We employed an optimized DGF basis set in order to obtain well-behaving
total wave functions at the triangle inequality boundaries (for
further details see Ref. \cite{isa04}).
The Gaussian functions are distributed equidistantly along the atom-atom
coordinates starting from 2.55 a$_{0}$ and out to 9.18 a$_{0}$ with a step of
0.17 a$_{0}$, a choice which ensures converged results within about 0.1 cm$^{-1}$ (0.01 \%
of the total ground state energy). 
In Table II we report the results obtained with DMC and DGF methods:
they are seen to be in good agreement with each other. Due to the addition of
the second light He atom, the Zero Point Energy (ZPE) of the trimer is slightly 
higher than the ZPE of the isolated dimer ion (see Table I). However, the two
ZPE values are very similar in percentage values, showing that the Li$^{+}$ impurity strongly affects
the features of the cluster, as expected from the involved potentials
while the additional helium atom has little effect on the bonding
features. The ionic system, as expected, does not present the typical
high degree of delocalization shown in pure He clusters
(the ZPE for He$_{3}$ is more than 99 \% of its total well depth \cite{10}), or by doped 
He aggregates with weakly bound impurities as, {\sl e.g.}, H$^{-}$, for which the 
impurity is clearly located outside the cluster (the ZPE for H$^{-}$He$_{2}$ 
is 90.66 \% \cite{nostroHeH}). Hence, we expect the Li$^{+}$-He$_{2}$ trimer (and
its larger clusters) to behave in a different way with respect to the
more weakly bound neutral He clusters
and thus presume that the classical description of the ionic
structures should give us realistic indications on the structure and 
energetics of such systems (as it was the case in, {\sl e.g.}, the H$^{-}$Ar$_{n}$ clusters 
that we studied earlier \cite{nostroArH}).

We therefore begin by looking at the average values of the radial distances and angles (see
Table II) which, together with the corresponding standard deviations, describe the ground 
state of the trimer and clearly confirm its having a rather floppy
structure: on the other hand, the Li$^{+}$ species still is undoubtedly
seen by our calculations to coordinate the two He atoms at a distance
determined  within 0.3 a$_{0}$. Information
on the overall geometrical features of the trimer is further gained from the radial and angular 
distributions shown in Fig. 3, where the values obtained with the classical optimization 
are also reported as vertical lines. DMC and DGF results are substantially concident and the small 
differences are mainly due to the bias contained in our DMC distributions.
We notice that the floppiness of the system is particularly evident when looking at the distribution function related
to the He-He distance, whose standard deviation is more than twice larger than the one
for the Li$^{+}$-He bonds. The classical values obtained in the
structure optimization are also closer to the quantum average values
obtained for the Li$^{+}$-He distances.

In the classical description we do find that the ground state of the
trimer is depicted by an isosceles triangle,
with the two shorter sides associated to the two Li$^{+}$-He distances and a longer one 
corresponding to the He-He distance. On the other hand, the real quantum system 
cannot be described by one single structure only, and its distribution
functions correctly show a delocalized triatom  with a dominant
contribution from the collinear arrangement. The image of a structure with the Li$^{+}$
coordinating the two He atoms at a rather well defined distance (notice the compact distribution
function related to Li$^{+}$-He distances) is not completely lost in the quantum description,  
meaning that the presence of the strong ionic forces from  Li$^{+}$ is reducing 
the degree of delocalization which is always present in the
pure He aggregates. This change determines the more rigid structure of the system in the
sense that now the  He atoms are
more strongly coordinated directly to the Li$^{+}$ impurity (see next section on the larger clusters). 

We carried out an additional analysis of the structural features of the trimer
by taking advantage of  the DGF pseudo-weights \cite{11}, which allow us to pictorially describe the
trimer in terms of types of triangular arrangements. In the upper
panel of Fig. 4 we thus report all the
employed basis set functions, grouped according to the triangular
family to which they belong, and in the lower panel of the same figure
we report the `weight' of the dominant families when describing
the ground state of the trimer. We can thus easily identify the most important
arrangement to be given by the `flat' isosceles, the collinear (with the the impurity in the middle)
and the scalene triangles, while all  other possibilities do not contribute in a
significant way. Again, we find that we cannot associate the system
to one unique structure. while 
the marked delocalization features are now mainly related  to the
He-He binding and less to the Li$^{+}$-He ionic forces; hence we see
that a conventional structure  with the Li$^{+}$ coordinating the two
He atoms can still be qualitatively identified. In the next section we shall
further discuss how the situation evolves with the addition
of more He atoms and to which extent the classical
results can be still seen to qualitatively correspond to the 
quantum description of their structures surrounding the ionic dopant.
     
\section{Energetics and structures of 
$\rm{\mathbf{Li^{+}\hspace{-0.11truecm}-\hspace{-0.11truecm}He_n}}$ 
(3 $\rm{\mathbf{\le n \le}}$ 30) 
clusters from quantum and classical calculations}

We start now to discuss the energetics and the geometrical features of the larger
$\mathrm{Li^+He}_n$ clusters with $n\ge3$. Up to
$n=10$, infact, it was still
possible to also carry out quantum DMC calculations which are not too demanding in terms
of CPU-time. Hence, for clusters of such a size we can make a direct comparison
between our quantum findings and the classical optimization results we obtained via the
combination of the OPTIM procedure \cite{walesOPT} with a random search for the minimum energy
structures (see for details refs. \cite{nostroArH,nostroNeH}). 
In Table III we report the results for the energetics. The left part
of that table shows the
minimum potential energies obtained by means of the classical optimizations 
(column labeled 'classical') in comparison with the corresponding DMC ground state 
energies (column  labelled 'quantum'). The differences between the two sets of values are due
to the ZPE effects of the nuclear motions: in the third column we also
display the ZPE value for each cluster as a percentage of the well depth. 
The amount of the ZPE effects increases as the cluster grows, since
the incresing addition of He atoms brings the ZPE percentage from about 20\% for
$\mathrm{Li^+He_2}$ to more than 40\% for the last cluster studied
here with the quantum DMC method ($\mathrm{Li^+He_{20}}$). 
On the right part of Table III we report the total energies relative to the loss of one He atom
between the pairs of $\mathrm{Li^+He}_n$ and $\mathrm{Li^+He}_{n-1}$ clusters, 
calculated both with the classical and quantum methods. We notice that the evaporation
energy is drastically reduced when passing from the cluster with n=6 to that with n=7 and
correspondingly the ZPE percentage value increases most markedly (more than 3\%). We
can then surmise that the structure with n=6 is a particularly stable
cluster as we shall further discuss below.

The data presented in Table III are pictorially reported in the two panels of Fig. 5 where
the energies are plotted as functions of the number n of He atoms in each cluster.
From the lower panel of Fig. 5 we can see the similar behaviour shown
by the $\Delta E_{ev}$ values up to
$n=6$ while for $7 \le n \le10$ both classical and quantum effects 
make the two curves show a marked drop in values. The single He atom evaporation energies, 
$\Delta E_{ev}=-\left[E(\mathrm{Li^+He}_n)-E(\mathrm{Li^+He}_{n-1})\right]$, 
(filled-in circles for the classical calculations 
and open square for those obtained with the DMC method) are plotted up
to $n=30$ (DMC resulòts up to $n=20$):
again the two curves show very similar behaviour, both presenting the same  
abrupt energy drop at $n=7$ and $n=9$. In contrast with
what we found in our analysis of H$^{-}$He$_{n}$ clusters \cite{nostroHeH}, the step-like 
structure shown by the classical treatment is now also present in the
quantum calculations.
Given such a correspondence between the classical and quantum description for these smaller
clusters it then becomes reasonable to try to explain the sudden
energy jumps of Fig. 5  by looking
at the lowest energy structures found with the classical 
minimizations (see Fig. 6). In that figure  we also report 
the corresponding symmetry groups, the total energy (in cm$^{-1}$), 
and the relevant distances between atoms (in a.u.). We therefore see
that the $\mathrm{Li^+He_6}$
has the  very symmetrical octahedral geometry (see second panel in the upper
part of the figure) where all the He atoms are equivalently coordinated to the central Li$^+$
at a distance very close
to the $\mathrm{R_{eq}}$ (3.58 a$_0$) of the $\mathrm{Li^+He}$ PEC. On
the other hand, when moving to the $\mathrm{Li^+He_7}$,
the repulsive forces acting between the rare gas species do not allow
any more for such a symmetrical arrangement around the $\mathrm{Li^+}$
ion and the net effect is that of decreasing the
energy contributions from the interactions between each He atom and the positive ion, i.e. the
$\mathrm{Li^+He}$ distance becomes larger, as one can see in Fig. 6  
from the reported $\mathrm{R_{LiHe}}$ values. Similar reasoning can be applied to explain 
the second energy step between $\mathrm{n=8}$ and $\mathrm{n=9}$: the evaporation energy
gives the mean value of the energy necessary to remove any of the  He atoms and the presence
of a non-equivalent rare gas atom in the apical position in $\mathrm{Li^+He_9}$ 
(see second panel in the lower part of Fig. 6) causes a significant decrease of the evaporation
energy with respect to its value for the $\mathrm{Li^+He_8}$ cluster.
Finally, for $\mathrm{Li^+He_{10}}$ the lowest energy structure we obtained with the classical
minimization corresponds to the symmetrical \emph{bicapped square antiprism} geometry.  
Hence, by using the classical geometries and energy minimization
procedures, we found three relatively more stable structures
for clusters with $n$=6,8 and 10, in correspondence with symmetrically
compact structures. However, we cannot
 associate the closing of a solvating `shell' to the clusters with $n$=6 and $n$=8 because they do not
constitute as yet a possible  core around which the larger clusters
grow.

The correspondence between classical and quantum structural pictures
is clearly well reproduced if we now
compare the quantum distribution functions with the classical results for the relative
distances and angles.   
In Fig. 7 we report the DMC atom-atom distribution functions for the 
$\mathrm{Li^+-He}$ and $\mathrm{He-He}$ distances within each $\mathrm{Li^+He}_n$ cluster, 
normalized to the total number n of 
possible 'connections' between $\mathrm{Li^+}$ and He atoms (solid lines) and to 
the total number $N=n(n-1)/2$ possible 'connections' between the n He atoms (dashed lines).
In this way we can make a direct comparison between the quantum
calculations and the lowest energy geometrical structures obtained with the classical
optimizations where the conventional picture of direct bonds existing
between localized, point-like, partners can be used. In that figure we
also report the classically optimized distances:
we have grouped together sets of close values and have given as horizontal bars their statistical
standard deviations: each set has a height proportional to the number of distances which have the same value. 
Finally, the displayed numbers are the values of the integration along
r for each broad peak 
in the quantum distribution functions and represent the number of
bonds within atoms which are in the given distance range under each
integration. 
In Fig. 7 we report the results for four selected 
clusters ($\mathrm{n=4,6,7,10}$) which we shall use in our discussion:
we also obtained similar results for all the other clusters.
When we look at the panel
showing the $\mathrm{Li^+He_4}$ clusters in the upper left of Figure 7, 
we see that the DMC distribution function for the $\mathrm{Li^+-He}$ distance peaks
at 3.88 a.u. while the classical distances are represented by one 'stick' at 3.59 a.u.
whose height is 4, equal to the value of the area under the quantum distribution (solid line); 
we see also that the DMC distribution function for the $\mathrm{He-He}$ distance 
peaks at 6.13 a.u. while the classical distances are represented by one 'stick' at 5.85 a.u. 
whose height is 6 (which is the value of $N=n(n-1)/2$, with  $n=4$) a number which is indeed
equal to the value of the area under the quantum distribution (dashed line).
For the larger clusters (see the other panels of the same figure) the number of distinct sets
of distances increases, but still the agreement between classical and quantum findings 
concerning the number of corresponding distances remains very good.
Although the mean values are different
as a consequence of the very diffuse behaviour of the wavefunctions in
such weak interatomic potentials, we can clearly see that the classical
results essentially provide the same qualitative structural picture.
In order to compare even more in detail the classical and quantum results,
we report in Fig 8 the DMC angular distribution functions P($\theta$)
for the angles centered at the Li$^+$ ion and at any of the He atoms
together with the results from classical optimization: we see again
that the agreement between
quantum and classical values (vertical lines) indeed remains very
close, at least at the qualitative level.

These findings allow us to make further comments on the microsolvation process
that occurs when the Li$^{+}$ is inserted in a small He cluster. Both the classical and the quantum
treatments concur in locating the Li$^{+}$ impurity inside the He$_{n}$ moiety as one can clearly
see from Figure 9  where we report the DMC distribution functions P(r) of the 
Li$^+$ (solid lines) and of the He atoms (dashed lines) from the geometrical center of 
each $\mathrm{Li^+He}_n$ cluster. 
This quantity is defined as:
\begin{equation}
\mathbf{r}_{gc}=\frac{1}{N} \sum_{i=1}^{N}\mathbf{r}_{i},
\end{equation}
where N runs over the total number of atoms in the cluster. The Li$^{+}$ is always closer to the
geometrical center with respect to the He atoms, and when the cluster
size increases the corresponding distribution functions have reduced
overlap, thereby showing that the cluster growth
is accompanied by the slow ``dropping'' of the Li$^{+}$ towards the
geometrical center of the latter

Finally, we carried out classical optimizations for larger clusters $\mathrm{Li^+He_n}$ (n=11-15,
18,20,22,26,30) in order to better confirm what we have observed in
the smaller ones; the lowest energy geometries for a selection of them are reported in Fig. 10.
For all the clusters under inspection, the growth occurs around the highly symmetric 
$\mathrm{Li^+He_{10}}$ core represented by 
the bicapped square antiprism polyhedron enclosing the Li$^+$ impurity
and drawn with thicker lines in the figure. 
The additional He atoms are now being placed further away
from the impurity without perturbing the structure of the $\mathrm{Li^+He_{10}}$ moiety 
which therefore seems to constitute the first solvation shell of the
atomic ion. We expect that the He atoms outside the shell
will be characterized by a greater delocalization and weaker interactions with the central Li$^{+}$
that the less shielded inner core of ten atoms. 
From now on we expect, therefore, that the cluster will grow by adding
more solvent atoms in a nearly isotropic fashion driven mainly by He-He interaction
and we surmise that the binding energies of each atom will become increasingly
similar to those of a pure He cluster. 
Correspondingly, the evaporation energy (see Fig. 6) now shows a markedly 
different behaviour with respect to what happens in the smaller clusters with n $\le$ 10:
the step-like feature disappears, to be substituted by a plateau giving us the average
energy needed to remove one of the nearly equivalent external He
atoms.

This expected result is also shown by the quantum calculations
(see Figure 5, lower panel) and is also borne out by the corresponding
quantum distributions of the helium adatoms given by the data of
Figure 11, where our DMC calculations for the
$\mathrm{Li}^+(\mathrm{He})_{20}$ clusters are reported: one clearly see
there that the radial distributions associated with the He distances from
the Li$^+$  moiety show a set of more compact values
related to the first shell of about 10 adatoms and a further
distribution at larger distances (and broader than the first one)
associated with the outer atoms that are chiefly bound by dispersion
and by strongly screened induction interactions.

\section{Conclusions}
In the present work we have analized the solvation process of a Li$^+$
ion in pure bosonic Helium clusters in order to  extend to a larger
number of atoms the studies we
had already carried out in previous work  on the smaller 
aggregates \cite{8}. Here a combination of classical energy minimization
techniques and of ``exact'' quantum Monte Carlo methods have been
employed in order to describe the structure and the energetics of the
Li$^+$(He)$_n$ clusters with $n$ up to 30 (20 for DMC calculations),
employing always a description based on the sum-of-potential
approximation.  

The basic approximation which this study relies on is that of 
calculating the full cluster interaction as a sum of pairwise
potentials. Although the error introduced by this assumption is in
general found to be (in absolute terms) larger in
ionic systems than in neutral ones, its relative weight
remains rather small \cite{8}. Thus, we believe that the resulting
stuctures and energies would not be substantially altered by the
correct use of a full Many-Body  potential (see for
example, the discussion in Ref. \cite{29} for the similar
situations of H$^-$ and Cl$^-$ in Rg clusters and our calculations in
Ref. \cite{8}). 

We have therefore shown that:
\begin{enumerate}
\item
  the Li$^+$ is fully solvated inside larger
  $^4$He clusters as already indicated by our earlier work on very
  small aggregates \cite{8};
\item
  the ionic Li$^+$ core does
  not form preferential ``molecular cores'' with surrounding He atoms but rather
  that the helium adatoms remain equivalently bound within each solvation
  shell to the central, solvated lithium ion that persists in carrying
  the positive charge for more than 98\% \cite{8}.
\item
  the ZPE corrections play a role in such systems, albeit strongly
  reduced with respect to the one found in neutral aggregates: this
  means that, at least in the initial solvation shell, the quantum
  adatoms are less delocalized and that the lithium-helium direct
  ``bonds'' are nearly rigid, classical  structures;
\item
  the classical optimization procedures can provide structural details
  which are reasonably close to those given by the quantum DMC
  calculations and can yield for ionic moieties the same structural
  picture as that given by the quantum treatment. 
\end{enumerate}
Furthermore, our comparison of single particle
evaporation energies given by classical and quantum results suggests
in both cases the formation of an initial shell of about ten He
atoms which are more strongly bound to the central ion. On the other
hand, beyond that initial shell the cluster growth appears to be
chiefly driven by He-He interactions, albeit at energies which are
initially still kept larger than those of the neutral systems by the
additional presence of the induction  field due to the
central ionic core. This bahevior may be compared to that of Na$^+$ and 
K$^+$ doped Helium clusters \cite{5a} where the first solvation shells were 
found to be of 9 and 12 atoms respectively. 

\begin{acknowledgments}
The financial support of the FIRB project, of the University of
Rome "La Sapienza"  Scientific Committee and of the European Union
"Cold Molecules" Collaborative Research Project no.
HPRN-CT-2002-00290 is gratefully acknowledged. One of us (M.Y.)
thanks I. T. U. Research Fund for the financial support and
I. T. U. High Performance Computing Center for the  computer time
provided.  We are grateful to Prof. M. Morosi and Dr. D. Bressanini for 
their help in improving the choice of trial functions. We also acknowledge 
the support of the INTAS grant 03-51-6170.
\end{acknowledgments}

\newpage
\begin{table}
  \caption{Bound states supported by the three pair potentials for
  Li$^+$-He. Values in the first column are taken from
  Ref. \cite{19}. The values in the second and third column have been
  calculated by us. All quantities are in units of cm$^{-1}$}
\begin{center}
\begin{tabular}{|c|c|c|c|}
\hline 
$ \nu$ & CCSD(T) (aug-cc-pV5Z) & MP4 (cc-pV5Z)& ATT \\
\hline
\hline 
0 & -519.975 & -515.225 & -496.319\\
1 & -311.396 & -307.280 & -289.972\\
2 & -165.739 & -161.536 & -150.728\\
3 & -75.323 &  -70.519 & -67.050\\
4 & -27.55 & -23.130 &  -23.902\\
5 & -7.191 & -4.642 & -5.905\\
6 & -0.925 & -0.356 & -0.669\\
7 & -0.007 & - & -0.002\\
\hline
\hline 
$D_{e}$ & 649.155 & 643.980 & 627.408\\
\hline
\hline 
ZPE & 129.180 (19.90 \%) & 128.755 (19.99 \%)& 131.090 (20.89 \%) \\
\hline
\end{tabular}
\end{center}
\end{table}

\clearpage
\newpage

\begin{table}
\begin{center}
\caption{Comparison of the results for the ground state of 
(LiHe$_{2}$)$^{+}$ obtained with both DMC and DGF methods. The energies are
in cm$^{-1}$, the distances in a.u., the angles in degrees and the areas S 
in a.u.$^{2}$. The ZPE is expressed as percentage of the the total well depth
for the trimer D$_{e}$ = 1306.02 cm$^{-1}$ (from classical optimization).}

\setlength{\tabcolsep}{0.13cm}
\begin{tabular}{|c|c|c|c|c|c|c|c|}
\hline
\hline
& E$_{v=0}$ & ZPE & r$_{He-He}$ & r$_{He-Li^{+}}$ & $\theta_{\widehat{He}}$ 
& $\theta_{\widehat{Li^{+}}}$ & $\sqrt{<S^{2}>}$ \\
\hline
\hline
DMC   & -1041.7 $\pm$ 0.68 & 20.24 \% & 6.59 $\pm$ 0.81 & 3.83 $\pm$ 0.32
& 28.43 $\pm$ 12.20 & 123.88 $\pm$ 24.01 & 5.54 $\pm$ 1.84 \\
\hline
DGF   & -1042.45 & 20.18 \% & 6.69 $\pm$ 0.73 & 3.75 $\pm$ 0.32 &  
26.94 $\pm$ 17.69 & 127.36 $\pm$ 27.99 & 5.24 \\
\hline
\hline
\end{tabular}
\end{center}
\end{table}

\clearpage
\newpage

\hoffset=0truein
\setlength{\tabcolsep}{0.23cm}
\begin{table}
\begin{center}
\caption{Minimum total energies and single-particle evaporation energies using classical and
quantum treatments for the nuclear motion. 
ZPE (\%) means: 
$\mathrm{\left[ \left(V^{TOT}_{classical}-V^{TOT}_{quantum}\right)/V^{TOT}_{classical}\right] \cdot 100}$.  
$\mathrm{\Delta E_{evap}}$ means: 
$\mathrm{-\left[V^{TOT}\left(Li^{+}(He)_{n}\right)-V^{TOT}\left(Li^{+}(He)_{n-1}\right)\right]}$
The quantum evaporation energy
for $n$=14 has been calculated using the
formula $E_{evap}(14)=(E(10)-E(14))/4$ and  for $n$=20  $E_{evap}(20)=(E(14)-E(20))/6$}. 
\vspace{0.8cm}
\begin{tabular}{|c|c|c|c||c|c|}
\hline 
\hline 
n & \multicolumn{3}{c||}{$\mathrm{V^{TOT}}$} & \multicolumn{2}{c|}{$\mathrm{\Delta E_{evap}}$} \\
\hline
& classical & quantum & ZPE (\%) & classical & quantum \\
\hline 
 2 & -1306.0 & -1041.7 $\pm$  0.7 & 20.2 & 656.8 & 521.1 $\pm$ 0.7 \\ 
 3 & -1970.5 & -1556.3 $\pm$  0.8 & 21.0 & 664.4 & 514.6 $\pm$ 1.5 \\
 4 & -2640.1 & -2038.1 $\pm$  2.9 & 22.8 & 669.6 & 481.7 $\pm$ 3.7 \\
 5 & -3287.0 & -2474.1 $\pm$  4.6 & 24.7 & 646.9 & 436.1 $\pm$ 7.5 \\
 6 & -3940.2 & -2895.3 $\pm$ 10.3 & 26.5 & 653.1 & 421.1 $\pm$ 14.9 \\
 7 & -4281.4 & -3000.7 $\pm$ 14.7 & 29.9 & 341.2 & 105.4 $\pm$ 25.0 \\
 8 & -4627.6 & -3161.5 $\pm$ 23.5 & 31.7 & 346.2 & 160.8 $\pm$ 38.2 \\
 9 & -4789.3 & -3234.3 $\pm$ 9.5  & 32.5 & 161.7 & 72.8  $\pm$ 33.0 \\
10 & -4936.0 & -3271.2 $\pm$ 19.5 & 33.7 & 146.7 & 36.9  $\pm$ 29.0 \\
14 & -5284.2 & -3292.4 $\pm$ 18.1 & 37.7 &  88.4 & 5.3   $\pm$ 37.6 \\
20 & -5785.6 & -3432.5 $\pm$ 56.1 & 40.7 & 72.2  & 23.3  $\pm$ 74.2 \\
\hline
\hline
\end{tabular}
\end{center}
\end{table}

\begin{figure}
  \includegraphics[width=1.0\textwidth]{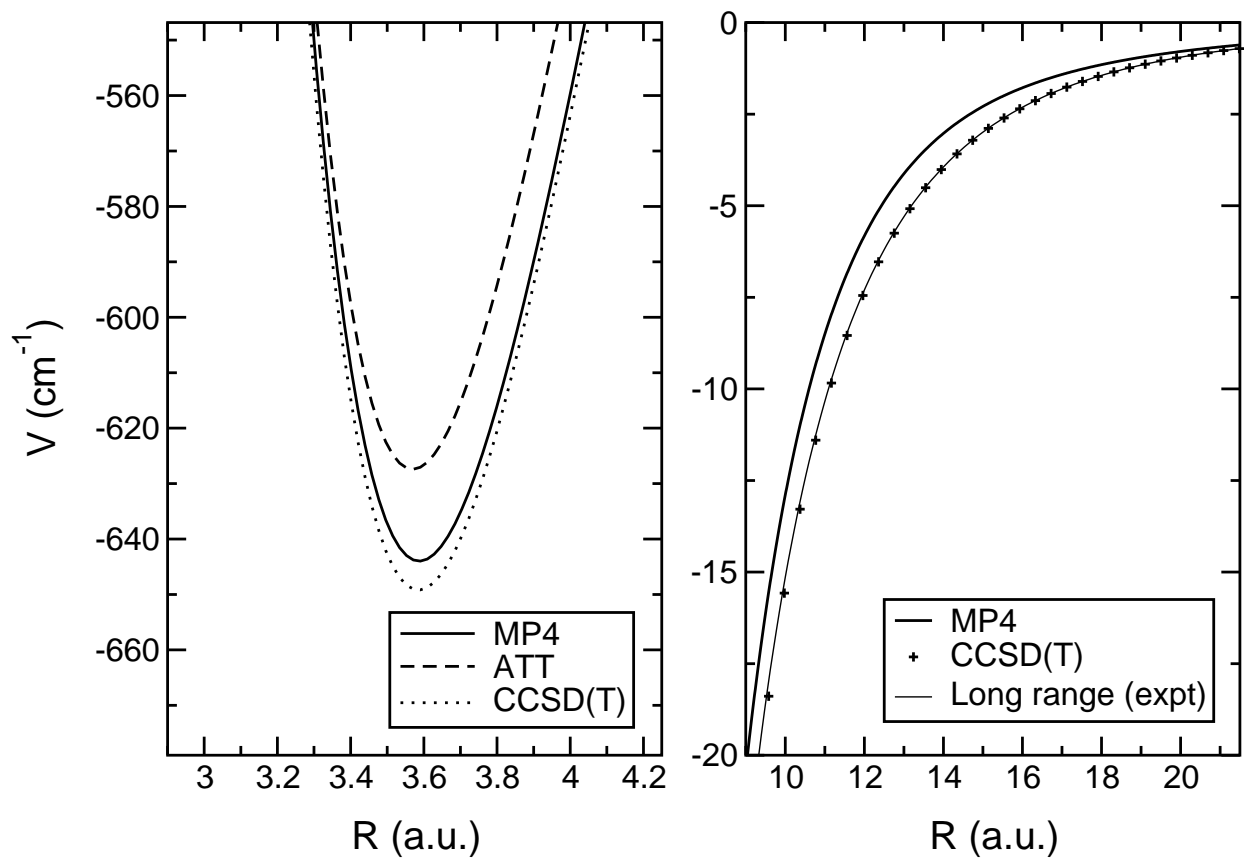}
  \caption{Left panel: comparison of different calculations of the well region
  Solid line: results from Ref. \cite{8}; dashes from
  ref.  \cite{17} (ATT); dots from Ref. \cite{19} (CCSD(T)). Right
  panel: comparison of the long-range polarization potential (thin
  line) with the long range behavior of two of the computed PEC's.}
\end{figure}

\begin{figure}
  \includegraphics[width=1.0\textwidth]{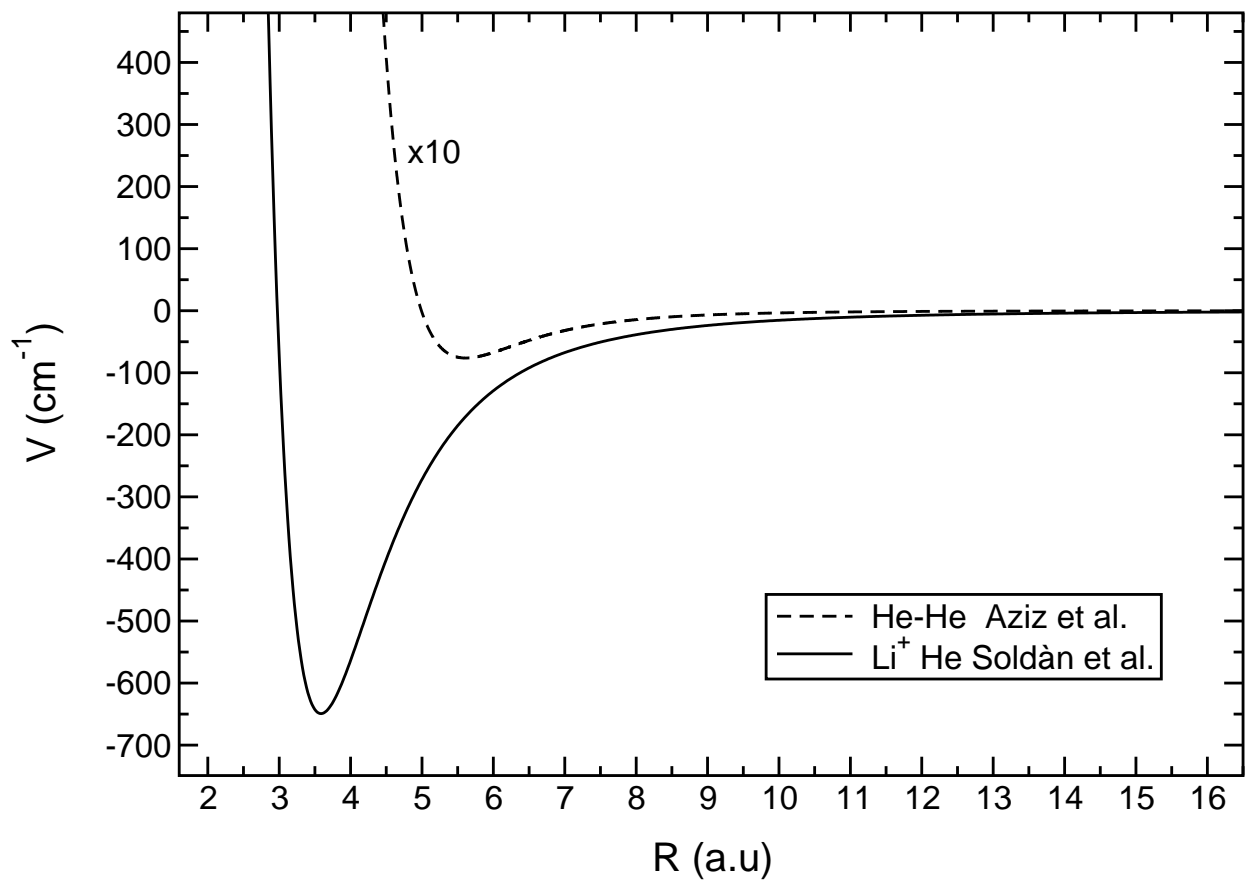}
  \caption{Comparison of the LiHe$^+$ potential from Ref. \cite{19}
  and the He-He one from Ref. \cite{24}. The He-He potential has been
  multiplied by ten.}
\end{figure}
  
\begin{figure}
  \includegraphics[scale=0.7]{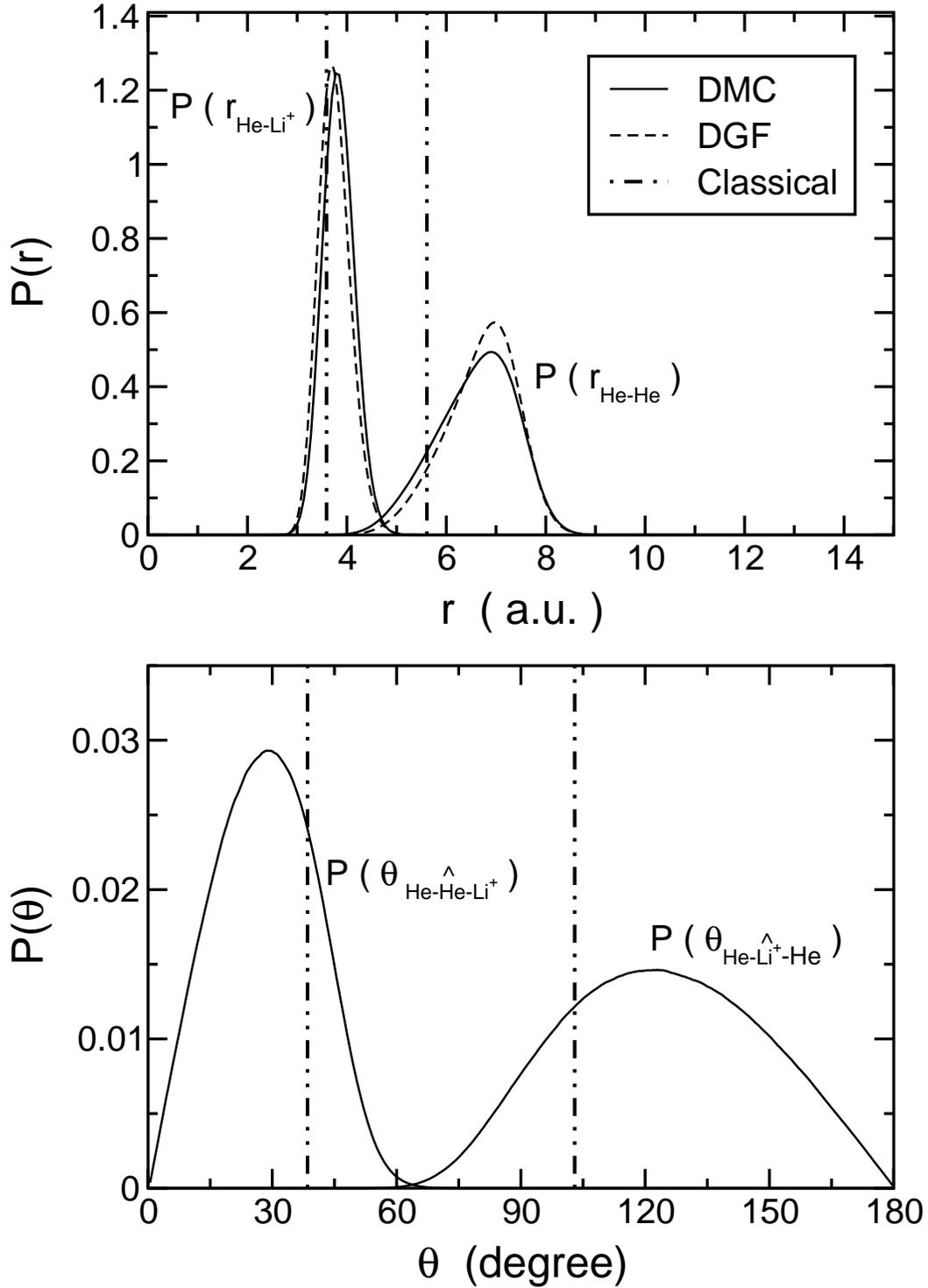}
  \caption{The distribution functions for the Li$^{+}$He$_{2}$
  calculated via the DMC (solid line) and the DGF (dashed line)
  methods. In the upper panel we report the pair distribution
  functions of the relative atom-atom distances while the
  lower panel shows the distribution functions for the internal
  angles. The (r,$\theta$) values obtained classically are given
  as vertical lines in the two panels.}
\end{figure}
  
\begin{figure}
  \includegraphics[width=1.0\textwidth]{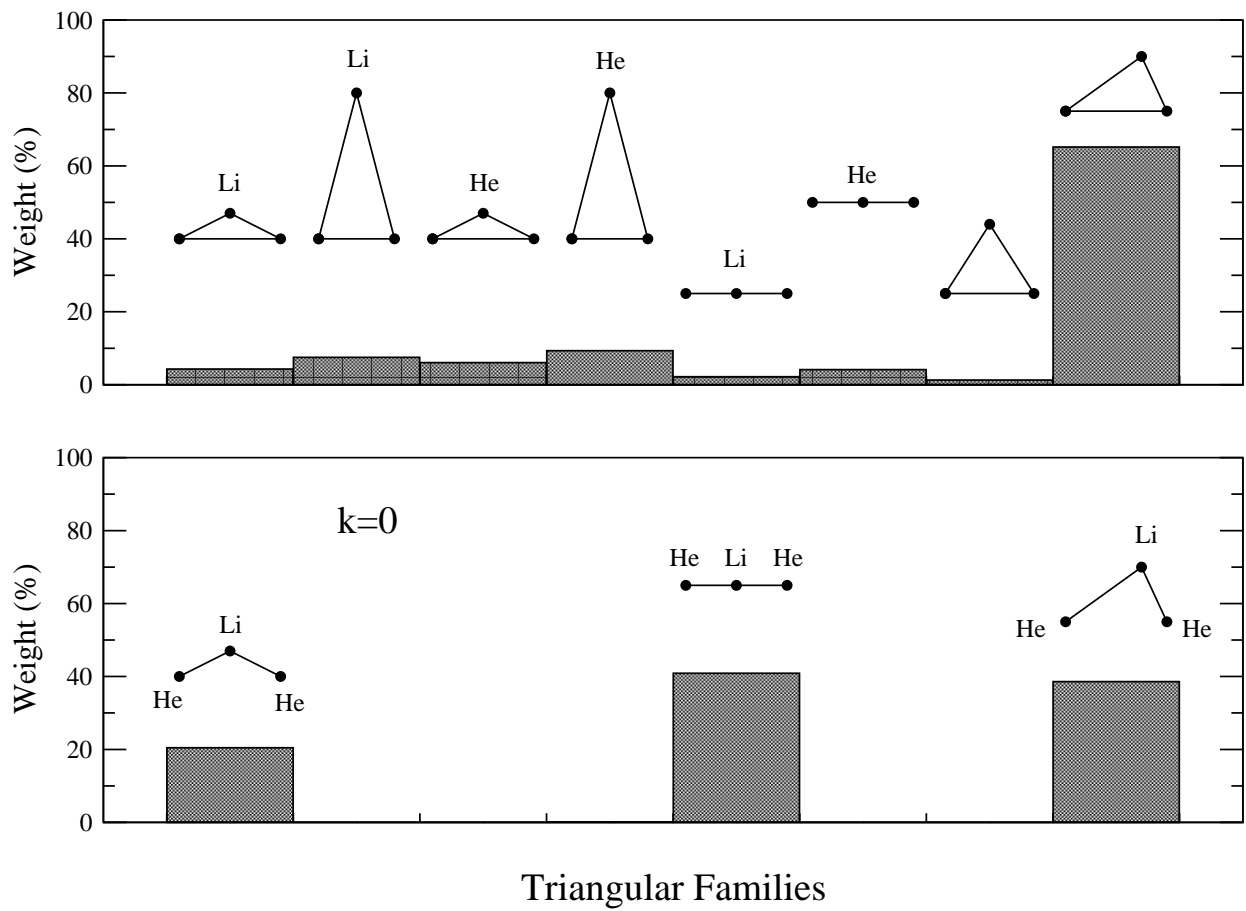}
  \caption{DGF description of the ground state of the Li$^{+}$He$_{2}$
  in terms of triangular families. In the upper panel we report those
  basis functions (as percentage with respect to the total number of basis
  functions) corresponding to the different types of triangular 
  configurations sketched in that panel.
  In the lower panel we show the three configurations that contribute
  the most to the ground state of the trimer.}
\end{figure}
  
\begin{figure}
  \includegraphics[height=0.8\textheight]{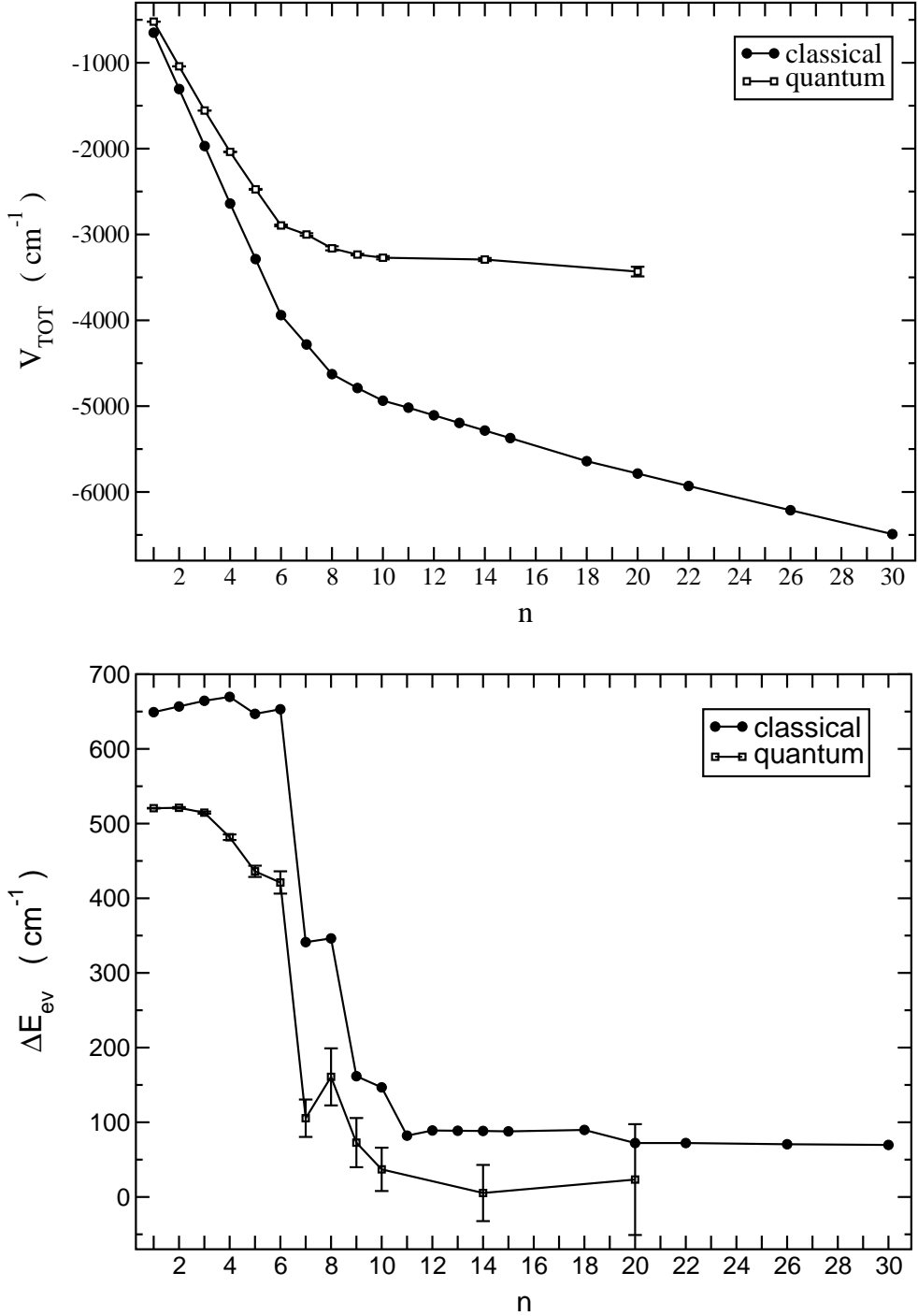}
  \caption{Upper panel: total energies (in cm$^{-1}$) as a function of the number n 
of helium atoms in $\mathrm{Li^+H_n}$ from classical optimizations 
(filled-in circles). Lower panel: single helium atom evaporation energies, 
$\Delta \mathrm{E=-\left[E(Li^+He_n)-E(Li^+He_{n-1})\right]}$
in $\mathrm{cm}^{-1}$, as a function of the number n of helium atoms in
$\mathrm{Li^+He_n}$ from the classical optimizations (filled-in
circles). The quantum results are given in both panels in open squares
  showing their corresponding numerical level of confidence.} 
\end{figure}

\begin{figure}
  \includegraphics[scale=0.7]{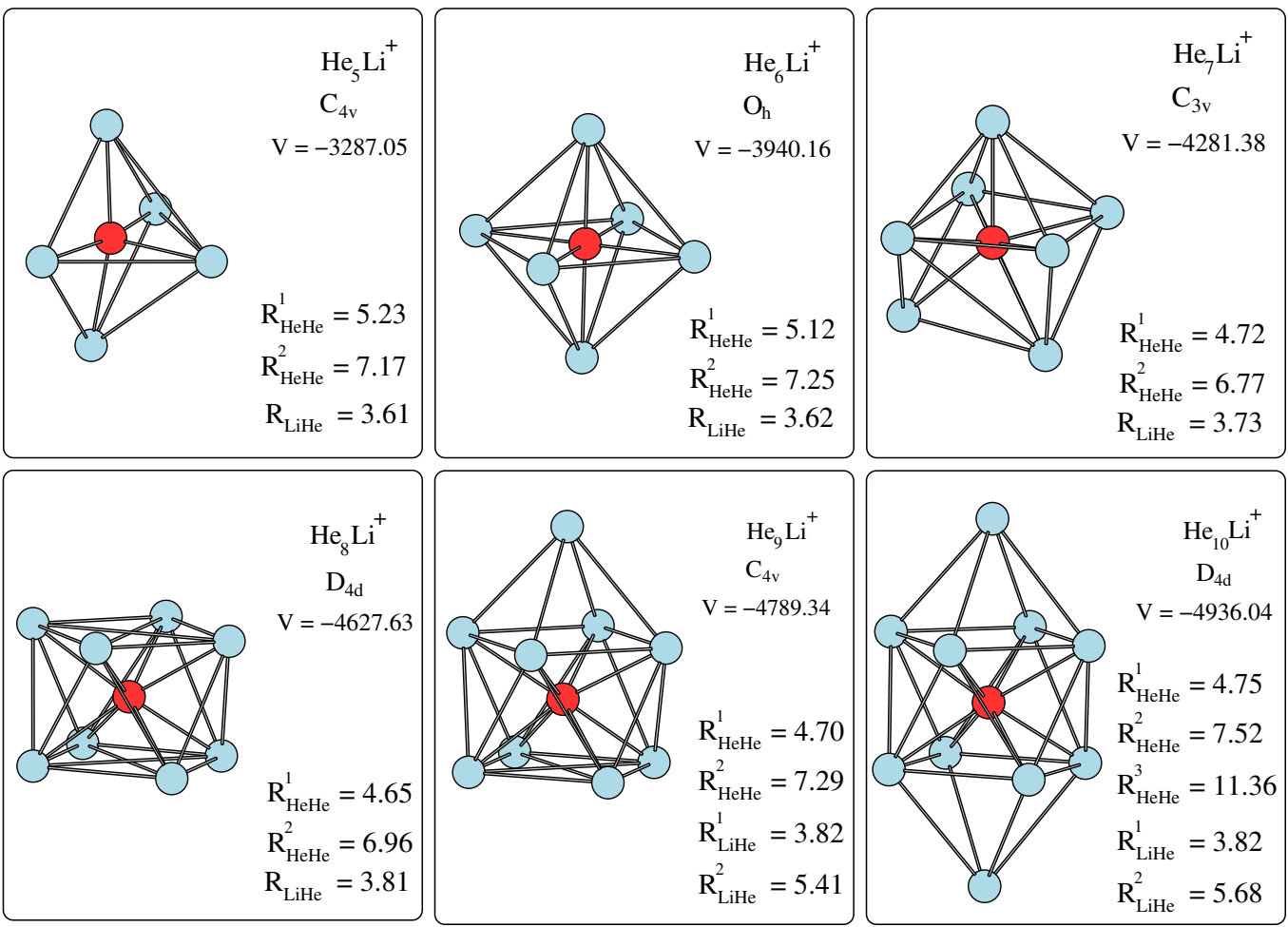}
  \caption{Optimized lowest energy structures obtained with the classical minimizations
for $\mathrm{Li^+He_n}$ with n from 5 to 10.}
\end{figure}

\begin{figure}
  \includegraphics[scale=0.5,angle=90]{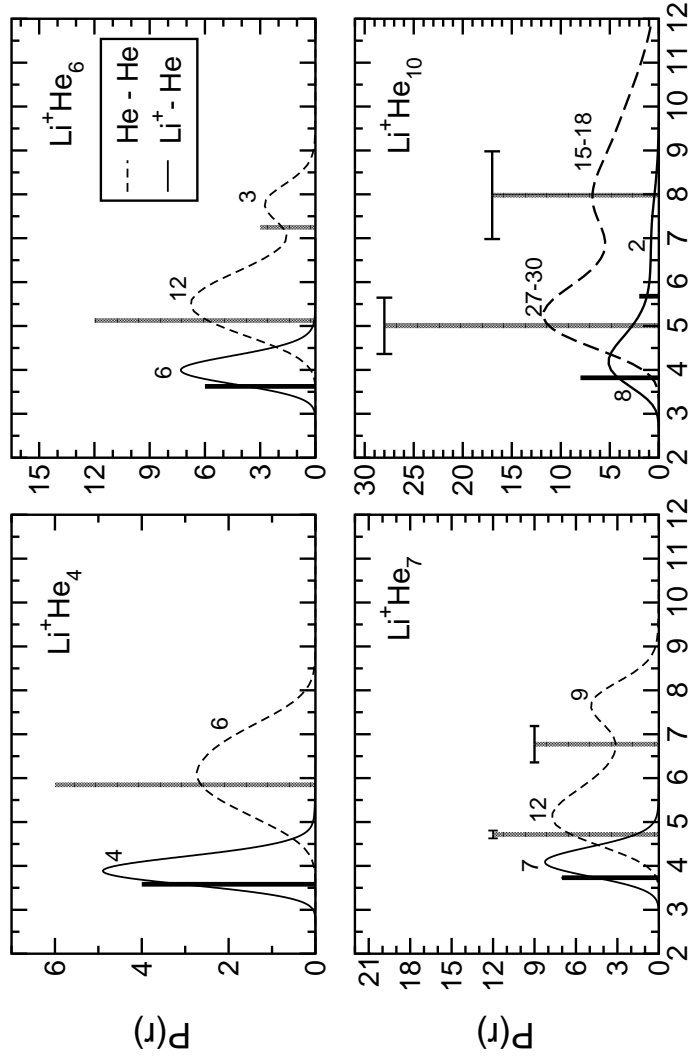}
  \caption{Distribution functions of the atom-atom relative distances P(r),
normalized to the number of 'connections' between the atoms
in each cluster, as a function of the distance r (in a.u.). 
DMC calculations for the $\mathrm{Li^+He_n}$, $\mathrm{n=4,6,7,10}$.
In all panels the solid lines are for the $\mathrm{Li^+-He}$ distances while the dashed
ones refer to the $\mathrm{He-He}$ distances. The vertical lines indicate the
computed mean values of the atom-atom distances (solid lines for
$\mathrm{Li^+-He}$  and gray lines for $\mathrm{He-He}$) of the optimized structures
from the classical calculations, with their height showing the number
of bonds associated with that distance (marked also by its spread
value on top of each ``stick''). The presented numbers in each panel
give the corresponding quantum  density of bonds within each
computed profile. }
\end{figure}

\begin{figure}
  \includegraphics[scale=0.65,angle=90]{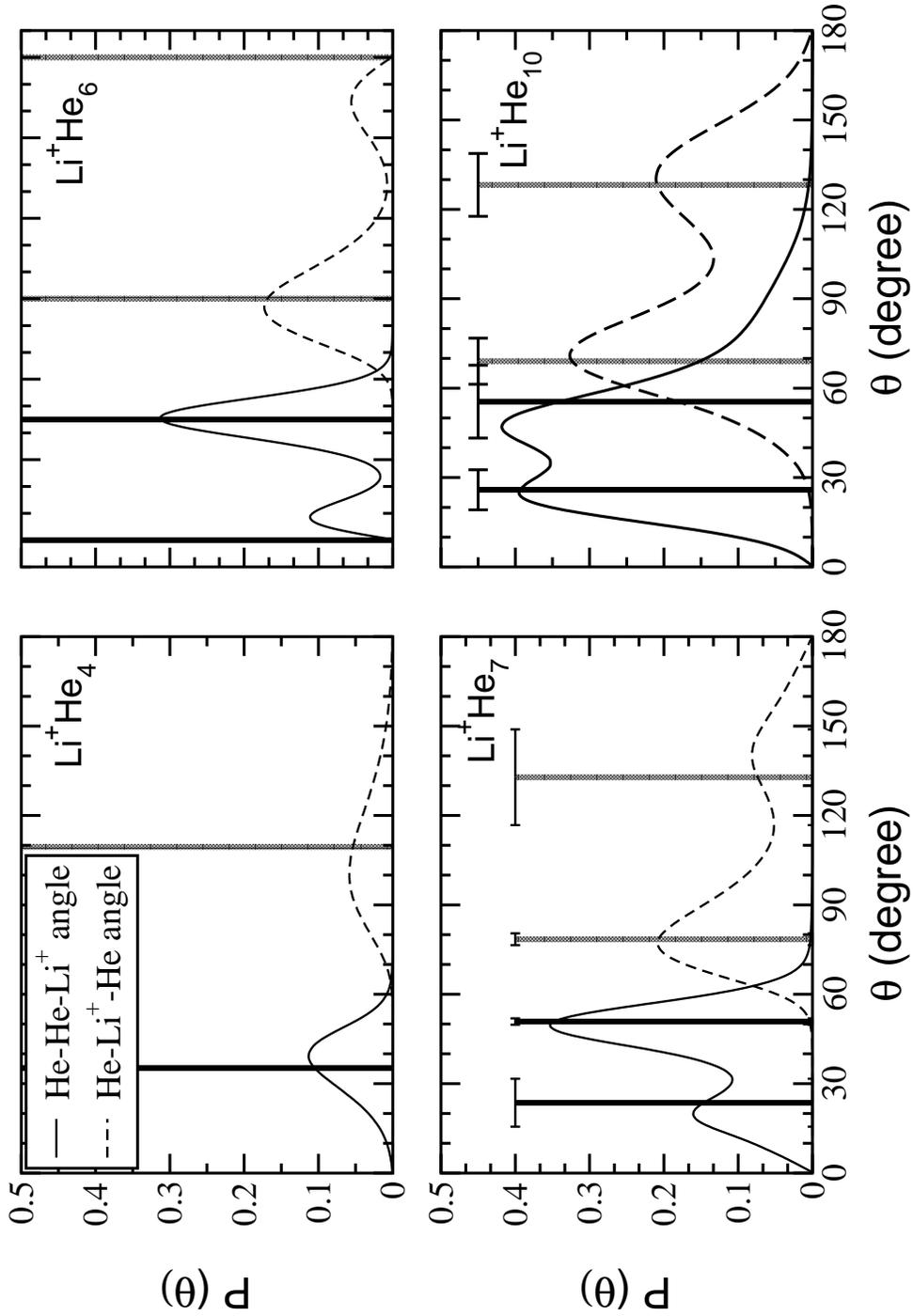}
  \caption{Angular distribution functions $\mathrm{P(\theta)}$,
obtained
with the DMC calculations for the $\mathrm{Li^+He_n}$, $\mathrm{n=4,6,7,10}$.
In all the panels the solid line is for the $\mathrm{He-He-Li^+}$ angles and the dashed
line for the $\mathrm{He-Li^+-He}$ angles. 
The vertical lines indicate the
various mean values of the angles (solid lines for $\mathrm{He-He-Li^+}$ 
and dashed lines for $\mathrm{He-Li^+-He}$) of the optimized structures
given by the classical calculations. The notation is the same as in
Figure 7}
\end{figure}

\begin{figure}
  \includegraphics[scale=0.75,angle=90]{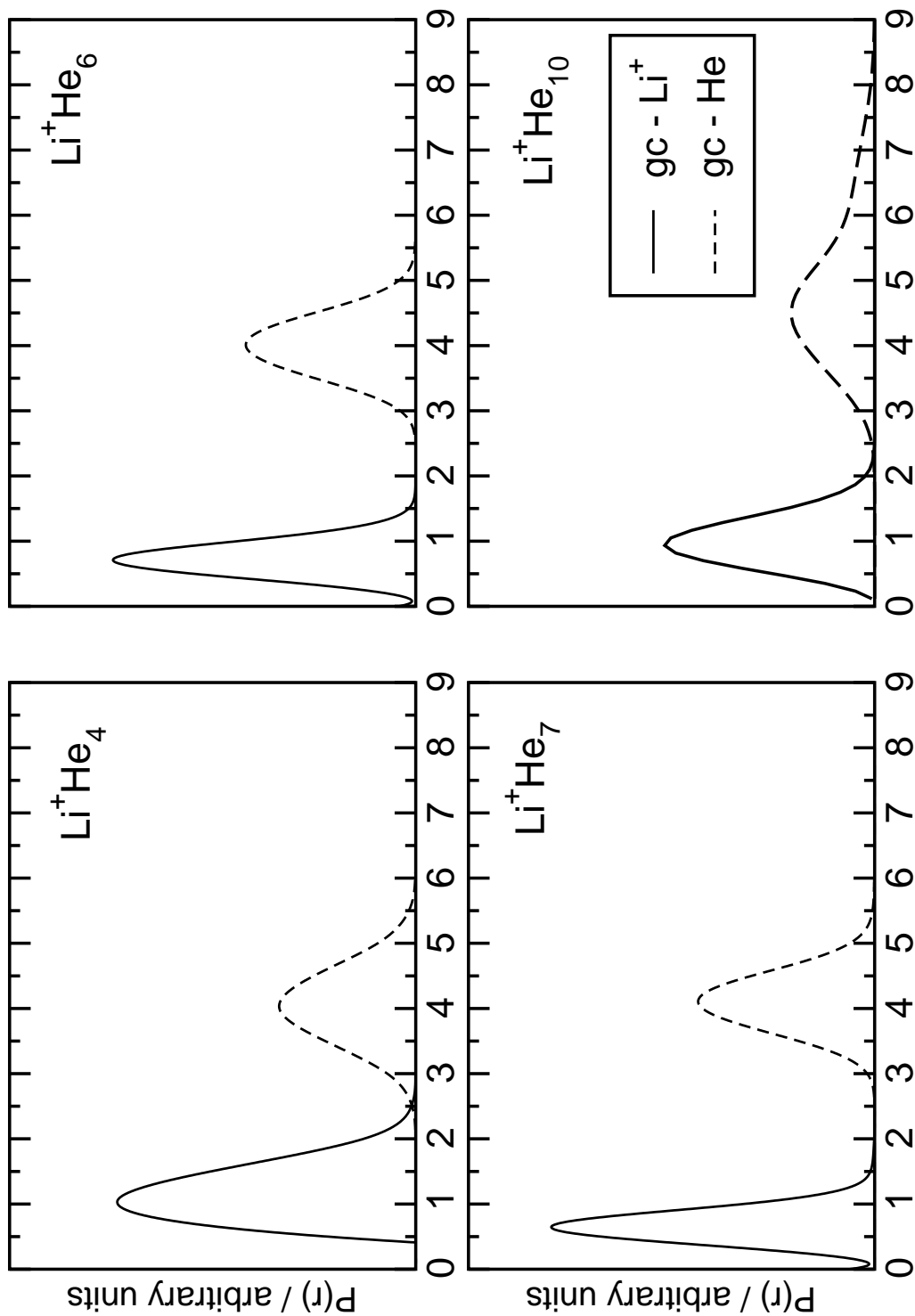}
  \caption{DMC distribution functions of the distances of the He atoms
(dashed lines) and of the Li$^{+}$ (solid lines) from the geometrical
    center (gc) for selected clusters as functions of the
distance r (in a.u.). The density distributions are normalized so that
the area under them is unitary}
\end{figure}

\begin{figure}
 \includegraphics[scale=0.80,angle=90]{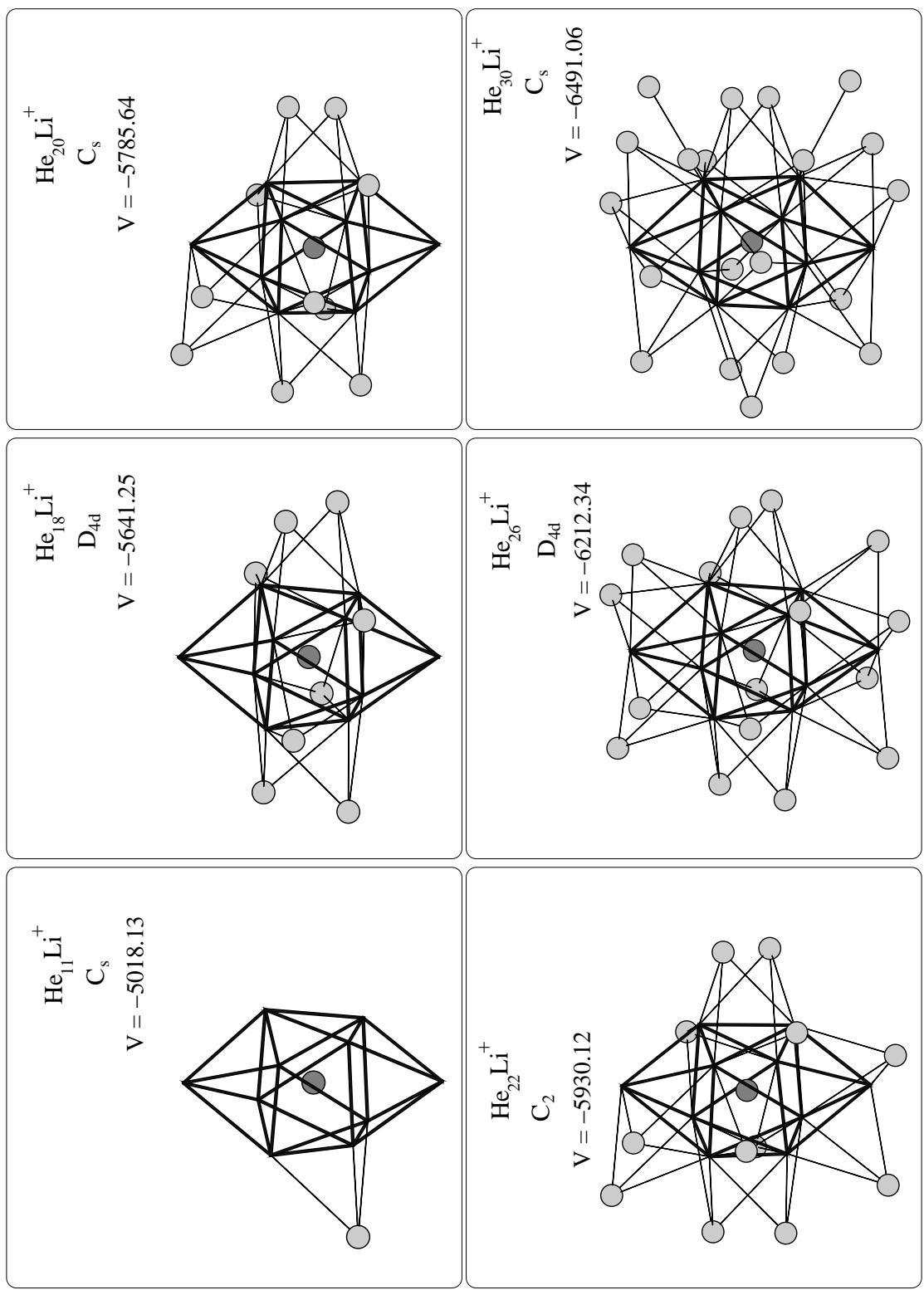}
  \caption{Optimized lowest energy structures obtained with the classical minimizations
for $\mathrm{Li^+He_n}$ with $\mathrm{n=11,18,22,26,30}$.
The central polyhedron including the Li$^+$ ion represents the $\mathrm{Li^+He_{10}}$ core.}
\end{figure}

\begin{figure}
 \includegraphics[width=0.9\textwidth]{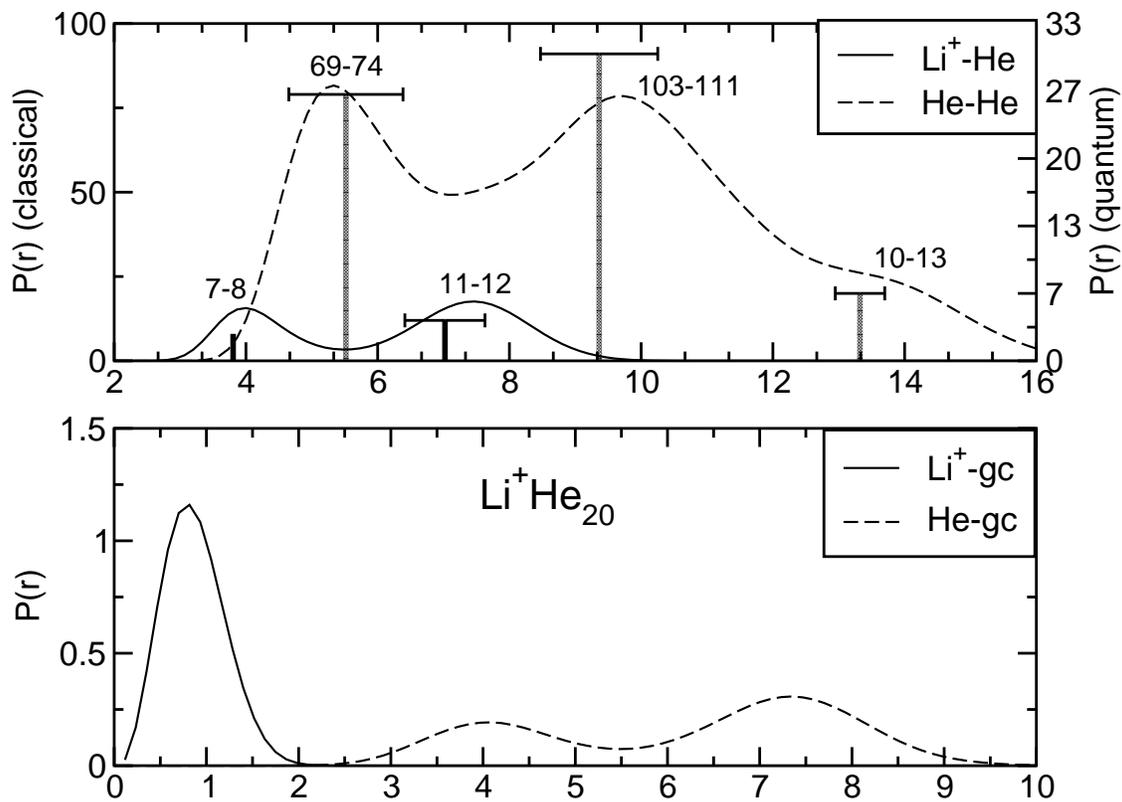}
  \caption{Quantum DMC distributions for the Li$^{+}$He$_{20}$ moiety. Upper panel: atom-atom 
           distance distributions compared with classical distance values as in Figure 8.
           Lower panel: distance distributions with respect to the geometric center of the cluster as 
           in Figure 9. }
\end{figure}

\end{document}